\documentclass[showpacs,preprintnumbers,amsmath,amssymb,twocolumn]{revtex4}
\usepackage{graphicx}% Include figure files
\usepackage{dcolumn}% Align table columns on decimal point
\usepackage{bm}% bold math
\begin{document}
\title{Modified Kubelka-Munk equations for localized waves inside a
  layered medium}

\author{Matthew M. Haney} \affiliation{Geophysics Department, Sandia
  National Laboratories, Albuquerque NM 87185-0750, USA}

\author{Kasper van Wijk} \affiliation{ Physical Acoustics Laboratory
  and Department of Geosciences, Boise State University, Boise ID
  83725, USA}

\begin{abstract}
  We present a pair of coupled partial differential equations to
  describe the evolution of the average total intensity and intensity
  flux of a wavefield inside a randomly layered medium. These
  equations represent a modification of the Kubelka-Munk equations, or
  radiative transfer. Our modification accounts for wave interference
  (e.g., localization), which is neglected in radiative transfer.  We
  numerically solve the modified Kubelka-Munk equations and compare
  the results to radiative transfer as well as to simulations of the
  wave equation with randomly located thin layers.
\end{abstract}


\maketitle

\section{Introduction}

The most basic mesoscopic theory that attempts to explain
multiply-scattered wave energy is the theory of radiative transfer
(RT). In a 1D layered medium, RT is equivalent to the well-known
Kubelka-Munk (KM) equations \cite{km,goedecke} because the KM
equations are in essence a two-flux theory and in 1D there are only
two directions (up and down).  However, due to the inevitability of
wave interference in 1D \cite{sheng}, RT is unable to accurately
predict all aspects of energy transport in randomly layered media
\cite{papa}.  Wave interference is explicitly ignored in RT
\cite{goedecke,loops} and leads to the phenomenon of wave
localization, as described by many authors previously
\cite{white,sheng,berry97,shapira05}. Wave localization is of primary
importance for topics such as the interaction of electrons with
disorder \cite{vollhardt} (e.g, the metal-to-insulator transition),
the transmission of light through randomly layered structures (such as
a stack of transparencies \cite{berry97}), and the late-time behavior
of seismic recordings at volcanoes \cite{wegler_loc}.

%Although wave interference is known to play an important role in 1D systems,
Previous studies have been devoted to understanding RT in layered
media, despite its neglect of wave interference.
%; however, the existing literature has not yet addressed the influence of wave interference on RT.
\citet{hemmer} may have been the first to solve for the Green's
function of RT in 1D, as pointed out by \citet{paasschens}. The
application of 1D RT to vertical seismic profiles has been the subject
of work by \citet{wu1,wu3} and \citet{wu2}. \citet{sato_fehler} have discussed 1D RT
and \citet{satoGJI} has derived the solution of the Green's function of RT
in 1D using the integral form, instead of the differential form used
by \citet{hemmer}. Building upon the work of \citet{wu2}, which
centered on stationary RT in layered media, the time dependent case
has been recently considered \citep{haney05}. \citet{bakut} have
generalized the Green's function of 1D RT in homogeneous media to the
case of a medium composed of piecewise homogeneous layers.  Though 1D
RT has been thoroughly understood in the course of these studies, how
wave interference changes the picture -- from the point of view of RT
-- has so far not been covered.  It is the aim of the present work to
properly incorporate wave interference, and hence the phenomenon of
wave localization, within the framework of RT for the case of layered
media.

Wave localization in 1D systems has received considerable attention,
both theoretically and experimentally. As a result, several different
techniques have been applied. Among the most widely used is random
matrix theory coupled with F\"{u}rstenberg's theorem
\cite{baluni,berry97,scales_vanvleck,mirkoGJI}. This approach deals
with the wavefield itself for a single realization of randomness by
using so-called ``self-averaging'' quantities \cite{shapiro_hubral}. Given an ensemble of
random realizations, these quantities converge (closely) to their
average in a single realization, provided the realization includes
enough scatterers.  In spite of its ability to model the wavefield
itself, random matrix theory is basically a stationary theory and it
relies on a limiting procedure, F\"{u}rstenberg's theorem, which takes
the limit as the number of matrix products (i.e., scatterers) becomes
infinite.
%A different method for studying wave localization 
Furthermore, random matrix theory is primarily limited to 1D systems.
Historically, this fact has led to a disconnect in the prevailing
theoretical treatment of multiple scattering in 1D (random matrix
theory) versus 2D and 3D (RT).

Significant progress has been made recently toward incorporating wave
interference into RT (at least within the diffusion approximation)
using the self-consistent (SC) theory of Anderson localization.  In
fact, a 1D version of the SC-theory has been studied analytically
\cite{lobkis:011112}.  %As alluded to above, t
The SC theory is different
from random matrix theory in that it predicts the late time spatial
and temporal evolution of the mean wavefield intensity (the squared
wavefield) for an ensemble of random realizations.
%In this way, it resembles RT (within the diffusion approximation) in that it makes predictions about the mean wavefield intensity;
%however, t
The crux of the SC theory is that it attempts to include the effects
of wave interference by using a so-called ``self-consistent''
expression for the diffusion constant, an idea originally popularized
by \citet{vollhardt}.

% we decsribe the phenomenon of localization from the point of view of radiative transfer? 
% wave localization had traditionally not been addressed from the point of view of radiative transfer theory
% however, in 1D Goedecke has provided a remarkably simple derivation of RT which explicitly highlights
% the omission of wave interference

Here, we attempt to include the effects of wave interference by
deriving the 1D RT equations from a fundamental level, using a
procedure first demonstrated by \citet{goedecke}.
%With this technique, interference terms appear explicitly in the derivation. By ignoring the 
%interference terms, the usual 1D RT equations
%take an alternate approach to
%include wave interference and thus localization into transport theory
%by examining the problem in layered media, i.e. the 1D case. 
We find that once properly modified, the 1D RT equations (also known
as the KM equations \cite{goedecke}) can account for interference
effects such as wave localization. We call these new equations
modified KM equations.  Thus, we are able to correctly account for
wave interference within the framework of RT, at least in 1D.  We also
show that the predictions of the modified KM equations agree with
predictions of random matrix theory, namely the expected exponential
decay of the steady-state transmission coefficient with sample size.
We finish by testing and verifying the modified KM equations through a
comparison with numerical simulations of the wave equation.  In
contrast to the 1D version of the SC theory \cite{lobkis:011112}, the
modified KM equations hold for all times and model both the total
intensity and the intensity flux.  At the end, we comment on the
prospects of generalizing the 1D theory to higher dimensions,
especially 3D where the notorious and interesting transition from
extended to localized wave propagation occurs.

\section{The scattering matrix with interference terms}

\begin{figure}
  \center \includegraphics[width=80mm]{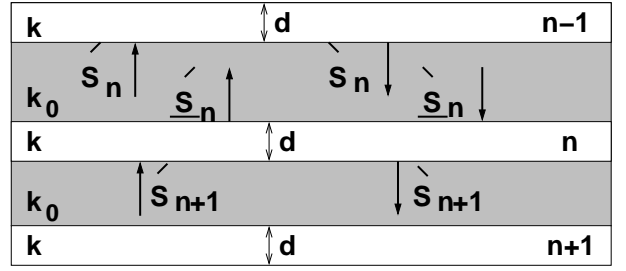}
  \caption{The up- and down-going waves near scatterer $n$.  We
    consider a random medium consisting of thin layers, or 1D
    scatterers, of thickness $d$ and local wavenumber $k$ embedded in
    a homogeneous background medium of wavenumber $k_{0}$.}
  \label{fig:upanddownatN}
\end{figure}

We aim to derive equations similar to the 1D RT equations, or KM
equations, but with the explicit inclusion of wave interference.
Although it is not necessary, we assume in the following that there is
no absorption for simplicity.  For a layered medium made up of thin
layers, or 1D scatterers, embedded in a homogeneous background medium,
the scattering matrix relating incident and scattered waves at
scatterer $n$ is
\begin{eqnarray}
&&\left[ \begin{array}{c}
       \acute{\underline{S}}_{n} \\
       \grave{S}_{n+1} \\
          \end{array} \right] = 
  \left[ \begin{array}{cc}
       ~~r~~& ~~t~~ \\
       ~~t~~& ~~r~~ \\
          \end{array} \right]~
  \left[ \begin{array}{c}
       \grave{\underline{S}}_{n} \\
       \acute{S}_{n+1} \\
          \end{array} \right],
\label{scatmat}
\end{eqnarray}
where $r$ and $t$ are the reflection and transmission coefficients of
a scatterer, $\grave{S}_{n+1}$ and $\acute{S}_{n+1}$ are the downward
and upward propagating complex wave amplitudes at the base of
scatterer $n$, and $\grave{\underline{S}}_{n}$ and
$\acute{\underline{S}}_{n}$ are the downward and upward propagating
complex wave amplitudes at the top of scatterer $n$, as shown in
Figure~\ref{fig:upanddownatN}.  Note that the complex wave amplitudes
at the top of scatterer $n$, $\grave{\underline{S}}_{n}$ and
$\acute{\underline{S}}_{n}$, are related to the complex wave
amplitudes at the base of scatterer $n-1$, $\grave{S}_{n}$ and
$\acute{S}_{n}$, by simple phase advance or delay
\begin{eqnarray}
\acute{\underline{S}}_{n}&=&{\acute{S}_{n} \over \sqrt{Z}} \nonumber \\
\grave{\underline{S}}_{n}&=&\sqrt{Z} \grave{S}_{n},
\label{adv_del}
\end{eqnarray}
where $\sqrt{Z}$ is the delay operator associated with the propagation
time between scatterers $n$ and $n-1$ \cite{fgdp}.  From equation~(\ref{adv_del}),
it follows that $|\acute{\underline{S}}_{n}|^2 = |\acute{S}_{n}|^2$
and $|\grave{\underline{S}}_{n}|^2 = |\grave{S}_{n}|^2$.  By taking
the squared magnitude of the two equations making up the scattering
matrix, equation~(\ref{scatmat}), and adding and subtracting them, we
thus arrive at the equations
\begin{eqnarray}
  &&
  \left| \acute{S}_n\right|^2 + \left| \grave{S}_{n+1}\right|^2 =
  \left( \left| r\right|^2+ \left| t\right|^2 \right)
  \left(  \left| \grave{S}_n\right|^2 +  \left| \acute{S}_{n+1}\right|^2 \right)
  + \nonumber \\
  &&
  \left( r^*t+ t^*r \right)
  \left( \grave{\underline{S}}_n^*\acute{S}_{n+1} +
    \grave{\underline{S}}_n\acute{S}_{n+1}^* \right),
    \label{eq:I}
\end{eqnarray}
\begin{eqnarray}
  &&
  \left| \acute{S}_n\right|^2 - \left| \grave{S}_{n+1}\right|^2 =
  \left( \left| r\right|^2- \left| t\right|^2 \right)
  \left(  \left| \grave{S}_n\right|^2 -  \left| \acute{S}_{n+1}\right|^2 \right)
  + \nonumber \\
  &&
  \left( r^*t- t^*r \right)
  \left( \grave{\underline{S}}_n^*\acute{S}_{n+1} -
    \grave{\underline{S}}_n\acute{S}_{n+1}^* \right).
    \label{eq:Iv2}
\end{eqnarray}
For a 1D scatterer embedded in a homogeneous medium, two identities
exist: $|r|^{2}+|t|^2=1$, or conservation of energy, and $r^*t+t^*r =
0$, as shown by \citet{ursin}.  With these identities,
expressions~(\ref{eq:I}) and (\ref{eq:Iv2}) can be simplified as
\begin{eqnarray}
  \left| \grave{S}_{n+1}\right|^2  -  \left| \acute{S}_{n+1}\right|^2 =\left| \grave{S}_n\right|^2 - \left| \acute{S}_n\right|^2,
  \label{eq:I_noxterms}
\end{eqnarray}
\begin{eqnarray}
  \left(\left| \grave{S}_{n+1}\right|^2  +  \left| \acute{S}_{n+1}\right|^2\right) -
  \left(\left| \grave{S}_n\right|^2 + \left| \acute{S}_n\right|^2\right) = \nonumber \\
  -2{|r|^2 \over |t|^2} \left(\left| \grave{S}_n\right|^2 - \left| \acute{S}_n\right|^2\right) 
  -4{\mbox{Im}(r^*t) \over |t|^2} \mbox{Im}(\grave{\underline{S}}_n\acute{S}_{n+1}^*).
  \label{eq:I_noxterms2}
\end{eqnarray}
Note that the interference terms are not present in
equation~(\ref{eq:I_noxterms}).  This equation states that the energy
flux between scatterers $n$ and $n-1$ equals that between scatterers
$n$ and $n+1$. The principle of energy flux conservation in layered
media has been used by Claerbout to derive the method of ``acoustic
daylight imaging''\cite[][in chapter 8]{fgdp}. 
In contrast, equation~(\ref{eq:I_noxterms2}), which describes the
local change in total intensity on either side of a scatterer,
contains an interference term.  This term depends on the correlation
between the downward propagating wavefield at the top of scatterer $n$
($\grave{\underline{S}}_n$) and the upward propagating wavefield at
the base of scatterer $n$ ($\acute{S}_{n+1}$).

We continue by averaging equations~(\ref{eq:I_noxterms}) and
(\ref{eq:I_noxterms2}) over ensembles of randomly placed scatterers.
We denote ensemble averages with brackets, $\langle\rangle$ and thus
the ensemble average of the squared magnitude of the down-going
wavefield between scatterers $n-1$ and $n$ by
$\langle|\grave{S}_n|^2\rangle = \grave{I}_n$ and so on for other
wavefield quantities.  With ensemble averaging, we obtain from
equations~(\ref{eq:I_noxterms}) and (\ref{eq:I_noxterms2}) that
\begin{eqnarray}
  \left(\grave{I}_{n+1}  -  \acute{I}_{n+1}\right) -
  \left(\grave{I}_n - \acute{I}_n\right) = 0,  
  \label{eq:I_noxtermsv2}
\end{eqnarray}
\begin{eqnarray}
  \left(\grave{I}_{n+1}  +  \acute{I}_{n+1}\right) -
  \left(\grave{I}_n + \acute{I}_n\right) = \nonumber \\
  -2{|r|^2 \over |t|^2} \left(\grave{I}_n - \acute{I}_n\right) %\nonumber \\
  -4{\mbox{Im}(r^*t) \over |t|^2} \mbox{Im}\langle\grave{\underline{S}}_n\acute{S}_{n+1}^*\rangle.
  \label{eq:I_noxtermsv22}
\end{eqnarray}
The stationary 1D RT equations result from these equations by first
assuming zero correlation in phase between wavefields propagating in
opposite directions at scatterer $n$,
$\langle\grave{\underline{S}}_n\acute{S}_{n+1}^*\rangle = 0$, followed
by taking a limiting procedure to move from discrete to continuous
variables, as shown by Goedecke \cite{goedecke}.

It is well known, however, that wave interference causes the term
$\langle\grave{\underline{S}}_n\acute{S}_{n+1}^*\rangle$ to be
nonzero, especially in 1D.  We thus include the term containing
$\langle\grave{\underline{S}}_n\acute{S}_{n+1}^*\rangle$, and
therefore account for interference effects, by considering the
directional wavefields on either side of a planar source within a 1D
random medium.  We take the depth axis ($z$-axis) positive downward.
First, consider the situation above the source depth $z_s$. There, the
up-going wavefield $\acute{\underline{S}}_{n}$ is the wavefield
incident from the source direction and it can be related to the
down-going wavefield $\grave{\underline{S}}_{n}$ using the reflection
coefficient $R_{1}$ for the entire random medium above scatterer $n$
as \cite{kennett}:
\begin{eqnarray}
  R_{1}\acute{\underline{S}}_{n} = \grave{\underline{S}}_{n}.
  \label{ref1}
\end{eqnarray} 
Moreover, the up-going wavefield $\acute{S}_{n+1}$ can be related to
the up-going wavefield on the other side of scatterer $n$,
$\acute{\underline{S}}_{n}$, using the reflection and transmission
coefficients of a single scatterer, $r$ and $t$, and the reflection
coefficient for the entire random medium below scatterer $n$, denoted
$R_{2}$, as:
\begin{eqnarray}
  \acute{\underline{S}}_{n} = {t \acute{S}_{n+1} \over 1 - r R_{2}}.
  \label{ref2}
\end{eqnarray}
Equations~(\ref{ref1}) and (\ref{ref2}) relate the wavefields
$\acute{S}_{n+1}$ and $\grave{\underline{S}}_{n}$:
\begin{eqnarray}
  \grave{\underline{S}}_{n} = {R_{1} t \acute{S}_{n+1} \over 1 - r R_{2}}.
  \label{ref3}
\end{eqnarray}
Substituting this relationship for $\grave{\underline{S}}_{n}$ into
equation~(\ref{eq:I_noxtermsv22}) and assuming the ensemble averaging
can be distributed as follows:
\begin{eqnarray}
  \mbox{Im}\langle{R_{1} t \over 1 - r R_{2}}|\acute{S}_{n+1}|^2\rangle = \langle|\acute{S}_{n+1}|^2\rangle
  \mbox{Im}\langle{R_{1} t \over 1 - r R_{2}}\rangle,
  \label{ref4}
\end{eqnarray}
equation~(\ref{eq:I_noxtermsv22}) above the source becomes
\begin{eqnarray}
  \left(\grave{I}_{n+1}  +  \acute{I}_{n+1}\right) -
  \left(\grave{I}_n + \acute{I}_n\right) = \nonumber \\
  -2{|r|^2 \over |t|^2} \left(\grave{I}_n - \acute{I}_n\right) %\nonumber \\
  -4\acute{I}_{n+1} {\mbox{Im}(r^*t) \over |t|^2} \mbox{Im} \langle{R_{1} t \over 1 - r R_{2}}\rangle.
  \label{eq:I_noxtermsv42}
\end{eqnarray}

Applying the same considerations to the situation below the source depth
$z_s$ means that the direction of the wavefield incident from the
source is the opposite of the case just shown. In addition, the roles
of the terms $R_{1}$ and $R_{2}$ are different: $R_{1}$ is now the
reflection coefficient for the entire random medium beneath scatterer
$n$ and $R_{2}$ is the reflection coefficient for the entire random
medium above scatterer $n$. This convention maintains the same
relation between $R_{1}$ and $R_{2}$ and the direction of the
incident wavefield as was used previously.  Thus, starting with
$R_{1}\grave{S}_{n+1} = \acute{S}_{n+1}$,
%where
%$R_{1}$ is now the reflection coefficient for the entire random medium
%beneath scatterer $n$, 
equation~(\ref{eq:I_noxtermsv22}) below the source is
\begin{eqnarray}
  \left(\grave{I}_{n+1}  +  \acute{I}_{n+1}\right) -
  \left(\grave{I}_n + \acute{I}_n\right) = \nonumber \\
  -2{|r|^2 \over |t|^2} \left(\grave{I}_n - \acute{I}_n\right) %\nonumber \\
  -4\grave{I}_{n} {\mbox{Im}(r^*t) \over |t|^2} \mbox{Im}\langle{R_{1}^{*} t^{*} \over 1 - r^{*} R_{2}^{*}}\rangle.
  \label{eq:I_noxtermsv52}
\end{eqnarray}
For a single realization of the ensemble, the $R_{1}$ and $R_{2}$ here
are not necessarily equal to the $R_{1}$ and $R_{2}$ considered
previously. However, the ensemble averages of the reflection
coefficients on either side of the source are the same since the
spacings of the scatterers above and below the source are drawn from the
same random distribution. From equations~(\ref{eq:I_noxtermsv42}) and
(\ref{eq:I_noxtermsv52}), the two situations differ not only by the
direction of the wavefield present in the last term ($\grave{I}_n$ or
$\acute{I}_{n+1}$), but also by a sign change in the last term (since
$\mbox{sgn}[\mbox{Im}\langle {R_{1}^{*} t^{*} \over 1 - r^{*}
  R_{2}^{*}}\rangle] = -\mbox{sgn}[\mbox{Im}\langle {R_{1} t \over 1
  - r R_{2}}\rangle]$).

\section{Modified KM Theory: the stationary case}
With these two cases (above and below the source), we now take the
limiting procedure -- as discussed by Goedecke \cite{goedecke} -- to
move from the discrete to the continuous case.  We examine here the
case below the source and then state the result for the case above the
source, since the procedure for the two cases is the same.  First,
note that $\acute{I}_{n+1}$ and $\grave{I}_{n+1}$ are defined at the
base of scatterer $n$, just as $\acute{I}_{n}$ and $\grave{I}_{n}$ are
defined at the base of scatterer $n-1$.  We define the average spacing
between the scatterers as $\rho^{-1}$ and thus the number of
scatterers per unit depth is $\rho$ (the number density).
%Thus, the average separation between the base of scatterer $n$ and the base of scatterer $n-1$ is 
Multiplying both sides of equation~(\ref{eq:I_noxtermsv52}) by $\rho$
results in
\begin{eqnarray}
  {\left(\grave{I}_{n+1}  +  \acute{I}_{n+1}\right) -
  \left(\grave{I}_n + \acute{I}_n\right) \over \rho^{-1}}= \nonumber \\
  -2\rho{|r|^2 \over |t|^2} \left(\grave{I}_n - \acute{I}_n\right) %\nonumber \\
  -4\rho\grave{I}_{n} {\mbox{Im}(r^*t) \over |t|^2} \mbox{Im}\langle{R_{1}^{*} t^{*} \over 1 - r^{*} R_{2}^{*}}\rangle.
  \label{eq:I_noxtermsv62}
\end{eqnarray}
%On average, the distance between the base of scatterer $n$ and the base of scatterer $n-1$ is $\rho^{-1}$.
As pointed out by Goedecke \cite{goedecke}, the term on the l.h.s. of
equation~(\ref{eq:I_noxtermsv62}) becomes a spatial derivative when
making the transition to a continuous depth variable $n\rho^{-1}
\rightarrow z$.
%under the continuum assumption, as discussed by Goedecke \cite{goedecke}.
Therefore, the directional wavefields become functions of $z$, that
is, $\grave{I}_{n} = I_{d}(z)$ and $\acute{I}_{n} = I_{u}(z)$ where
$I_{d}$ and $I_{u}$ are the ensemble-averaged down-going and up-going
intensities.  Therefore, equation~(\ref{eq:I_noxtermsv62}) becomes
\begin{eqnarray}
   {d \left(I_{d} + I_{u}\right) \over dz} = \nonumber \\
  -2\rho{|r|^2 \over |t|^2} \left(I_{d} - I_{u}\right) %\nonumber \\
  -4\rho I_{d} {\mbox{Im}(r^*t) \over |t|^2} \mbox{Im}\langle{R_{1}^{*} t^{*} \over 1 - r^{*} R_{2}^{*}}\rangle.
  \label{eq:I_noxtermsv72}
\end{eqnarray}
We further simplify equation~(\ref{eq:I_noxtermsv72}) by defining the
ensemble-averaged total intensity $I_{t}=I_{d}+I_{u}$ and the
ensemble-averaged intensity flux $I_{f}=I_{d}-I_{u}$.  This simplifies
equation~(\ref{eq:I_noxtermsv72}) as
\begin{eqnarray}
  {d I_{t} \over dz} = 
  -2\rho{|r|^2 \over |t|^2} I_{f} %\nonumber \\
  -4\rho I_{d} {\mbox{Im}(r^*t) \over |t|^2} \mbox{Im}\langle{R_{1}^{*} t^{*} \over 1 - r^{*} R_{2}^{*}}\rangle.
  \label{eq:I_noxtermsv82}
\end{eqnarray}
We finally define the scattering mean free path $\ell_{s}$, the
localization length $\ell_{loc}$, and a dimensionless parameter $B$ as
\begin{eqnarray}
  {B \over \ell_{s}}&=& \rho{|r|^2 \over |t|^2} \nonumber \\
  {1 \over \ell_{loc}}&=& 2\rho{\mbox{Im}(r^*t) \over |t|^2}\mbox{Im}\langle{R_{1}^{*} t^{*} \over 1 - r^{*} R_{2}^{*}}\rangle.
  \label{scatandloc}
\end{eqnarray}
Using these parameters, we can rewrite
equation~(\ref{eq:I_noxtermsv82}) concisely as
\begin{eqnarray}
  {d I_{t} \over dz} =
  {-2B \over \ell_{s}}I_{f} - {2 \over \ell_{loc}} I_{d}.
  \label{eq:I_noxtermsv92}
\end{eqnarray}

We have chosen the definitions in equation~(\ref{scatandloc}) in a
manner consistent with the usual definitions for these quantities
\cite{goedecke,sheng}. For instance, regarding the quantity $\rho
|r|^2 / |t|^2$, in the weak scattering limit ($|t|^2 \approx 1$) we
find that
%regarding the scattering mean free path, in the weak scattering limit, when $|t|^2 \approx 1$, we define $B$ according to
\begin{eqnarray}
  \rho{|r|^2 \over |t|^2} \approx \rho |r|^2  = \rho (|r|^2 + |t-1|^2) {|r|^2 \over |r|^2 + |t-1|^2} = {B \over \ell_{s}},
  \label{scat_demonstrate}
\end{eqnarray}
where
\begin{equation}
  B = {|r|^2 \over |r|^2 + |t-1|^2},
\end{equation}
Thus, $B$ is a dimensionless parameter
describing the directionality of the scattering \cite{haney05}. For
isotropic scattering, $B=1/2$. In addition we define 
\begin{equation}
  {1 \over \ell_{s}} = \rho (|r|^2 + |t-1|^2),
  \label{smfp}
\end{equation}
consistent with what we know for the 1D scattering cross section from
\citet{sheng}: $\sigma_{s} = |r|^2 + |t-1|^2$.  Therefore, from
equation~(\ref{smfp}), we can identify the factor $\rho (|r|^2 +
|t-1|^2) = \rho\sigma_{s}$.  In the weak scattering limit, it is well
known that $\ell_{s} = 1/\rho\sigma_{s}$. Thus, our definition for
$\ell_{s}$ in equation~(\ref{smfp}) is consistent with the usual
definition of $\ell_{s}$ in the weak scattering limit.
Appendix~\ref{app:loca} demonstrates the consistency of the definition
for $\ell_{loc}$ as it appears in equation~(\ref{eq:I_noxtermsv92})
based on the relation in equation~(\ref{scatandloc}).

We have just shown how to apply the limiting procedure to
equation~(\ref{eq:I_noxtermsv52}).  Applying the same limiting
procedure to equations~(\ref{eq:I_noxtermsv2}) and
(\ref{eq:I_noxtermsv42}) gives all of the necessary stationary
transport equations, which we summarize here:
%, and (\ref{eq:I_noxtermsv52}), we obtain the
%following stationary transport equations:
\begin{equation}
  {d I_{f} \over dz} = 0,
  \label{ir_stat2}
\end{equation}
\begin{eqnarray}
  {d I_{t} \over dz}&=&
  {-2B \over \ell_{s}}I_{f} + {2 \over \ell_{loc}} I_{u}~~~~~\mbox{for $z < z_{s}$} \nonumber \\
   &=&
  {-2B \over \ell_{s}}I_{f} - {2 \over \ell_{loc}} I_{d}~~~~~\mbox{for $z > z_{s}$}.
  \label{il_stat2}
\end{eqnarray}
%In the above equations, $I_{t}$ is the ensemble-averaged total
%intensity, $I_{f}$ is the ensemble-averaged intensity flux, and
%$z_{s}$ is the source depth.  The relationships between $I_{t}$ and
%$I_{f}$ and the ensemble-averaged down-going and up-going intensities
%are $I_{d} = (I_{t}+I_{f})/2$ and $I_{u} = (I_{t}-I_{f})/2$.
%Additionally, $\ell_{s}$ is the scattering mean free path,
%$\ell_{loc}$ is the localization length, and $B$ is a parameter
%describing the directionality of the scattering \cite{haney05}. For
%isotropic scattering, $B=1/2$. We have grouped various factors
%appearing in equation~(\ref{eq:I_noxtermsv52}), for instance those
%containing $r$, $t$, $R_{1}$, and $R_{2}$, and renamed them $\ell_{s}$
%and $\ell_{loc}$.
%the scattering mean free path $\ell_{s}$
%and the localization length $\ell_{loc}$. 
%We have done so in a manner consistent with their usual definitions
%\cite{goedecke,sheng}.  For instance, we invoke $\ell_{loc}$ since it
%is known, from random matrix theory \cite{berry97} that in the
%stationary case $I_{t} \sim \mbox{exp}(-|z-z_{s}|/\ell_{loc})$, as the
%above equations also predict. 
These equations comprise the modified KM equations in the stationary
case.  Equation~(\ref{il_stat2}) may be rewritten more concisely as
\begin{equation}
  {d I_{t} \over dz} =
  -2\left[{B \over \ell_{s}}+{1 \over \ell_{loc}}\right]I_{f} - {\mbox{sgn}(z-z_{s}) \over \ell_{loc}} (I_{t} - |I_{f}|),
  \label{il_stat}
\end{equation}
where the quantity $I_{t} - |I_{f}|$ is either $2I_{u}$ for $z >
z_{s}$ or $2I_{d}$ for $z < z_{s}$: it is twice the intensity
propagating back toward the source.  Equation~(\ref{il_stat}) shows
that the inclusion of wave interference in the KM (or RT) equations
leads to two additional terms which affect the average total intensity
in different ways. The first term containing $1/\ell_{loc}$ on the
r.h.s.  of equation~(\ref{il_stat}) causes the coherent intensity to
decay more rapidly than when wave interference is neglected.
Furthermore, the second term containing $1/\ell_{loc}$ on the r.h.s.
causes the spatial distribution of the incoherent intensity to be
entirely different than in the the case of no interference (as
demonstrated later in a numerical example). The form of
equation~(\ref{il_stat}) allows the identification of the extinction
mean free path (the decay of the coherent intensity), $1/\ell_{ext} =
B/\ell_{s} + 1/\ell_{loc}$.  This insight is possible since the
quantity $I_{t} - |I_{f}|$ in equation~(\ref{il_stat}) is zero for the
coherent intensity.  Note that the coherent intensity decays
exponentially even without interference effects ($\ell_{loc}
\rightarrow \infty$, or RT) due to scattering out of the forward
direction.

% justify localization length  in appendix %

\section{Modified KM theory: the time-dependent case}

Having derived the modified KM equations for the stationary case in
equations~(\ref{il_stat}) and (\ref{ir_stat2}), we will turn our
attention to the time-dependent (dynamic) case.
%The above stationary equations can be easily generalized to the
%dynamic case, 
Given the current coordinate system for $z$, this is accomplished by
noting that $d I_{u}/dz = \partial I_{u}/\partial z + v^{-1} \partial
I_{u}/\partial t$ and $d I_{d}/dz = \partial I_{d}/\partial z - v^{-1}
\partial I_{d}/\partial t$, where $v$ is the velocity of energy
transport (the energy velocity). Including the presence of planar
isotropic (zero net down-going component) sources \cite{haney05}, we obtain the
following time-dependent equations
\begin{equation}
  {\partial I_{f} \over \partial z} + {1 \over v}
  {\partial I_{t} \over \partial t} =
  {\Gamma \over v},
  \label{ir}
\end{equation}
\begin{equation}
  {\partial I_{t} \over \partial z} + {1 \over v}
  {\partial I_{f} \over \partial t} =
  -2\left[{B \over \ell_{s}}+{1 \over \ell_{loc}}\right]I_{f} -
  {\mbox{sgn}(z-z_{s}) \over \ell_{loc}} (I_{t} - |I_{f}|),
  \label{il}
\end{equation}
%where $v$ is the velocity of energy transport (the energy velocity)
where $\Gamma$ is the isotropic (omnidirectional) source term \cite{haney05}.
Note that for $\ell_{loc}\rightarrow \infty$ (no wave interference),
equations~(\ref{ir}) and (\ref{il}) are the same equations as have
been studied previously by others within the context of RT in layered
media \cite{wu1,wu2,wu3,haney05}.

With equations~(\ref{ir}) and (\ref{il}), which are the modified KM
equations in the time-dependent case, we proceed to numerically solve
the equations for two cases: with interference and without
($\ell_{loc} \rightarrow \infty$, or RT). These cases are compared to
ensemble averages of simulations of the wave equation.

\begin{figure*}
  \includegraphics[width=7cm]{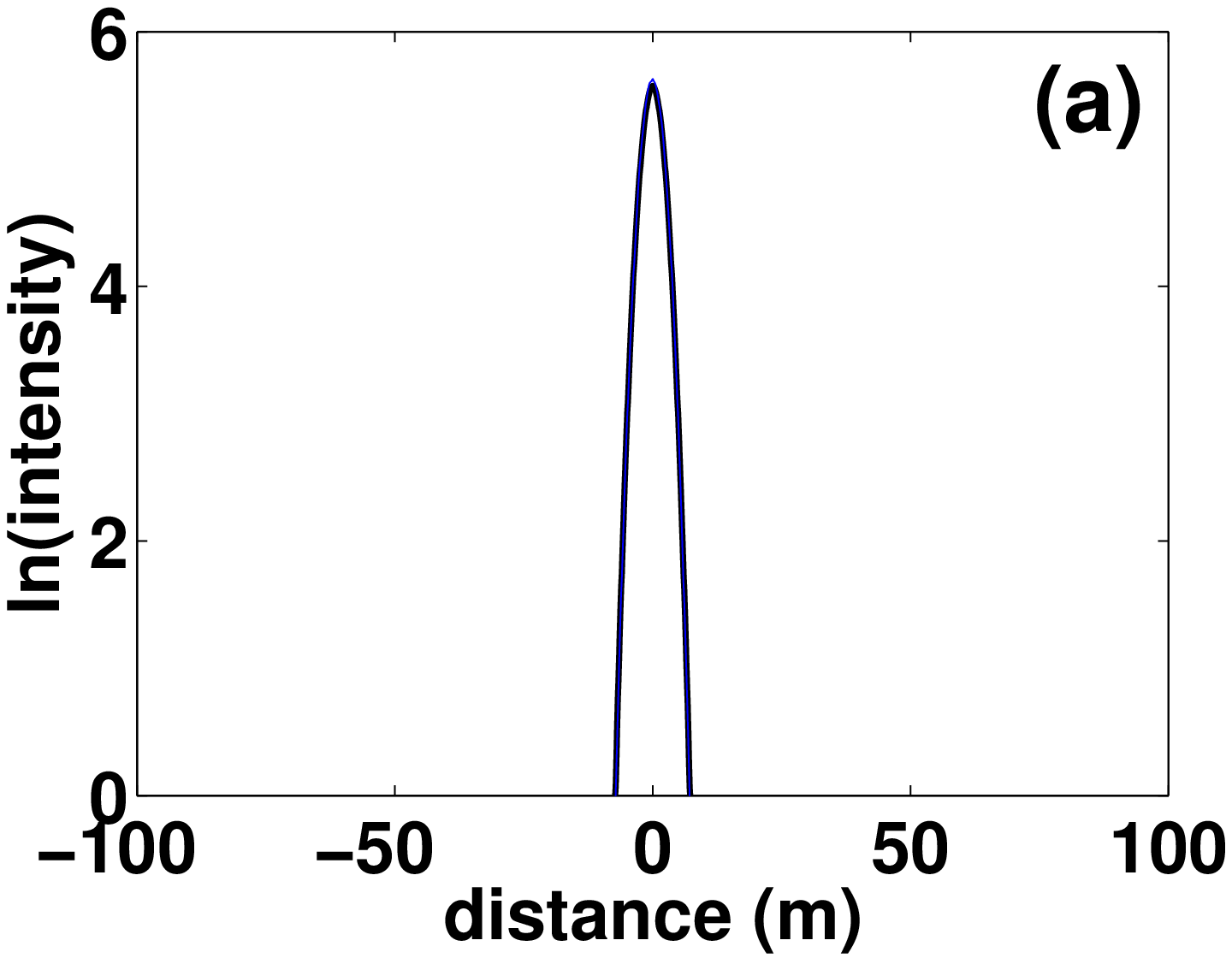}
  \includegraphics[width=7cm]{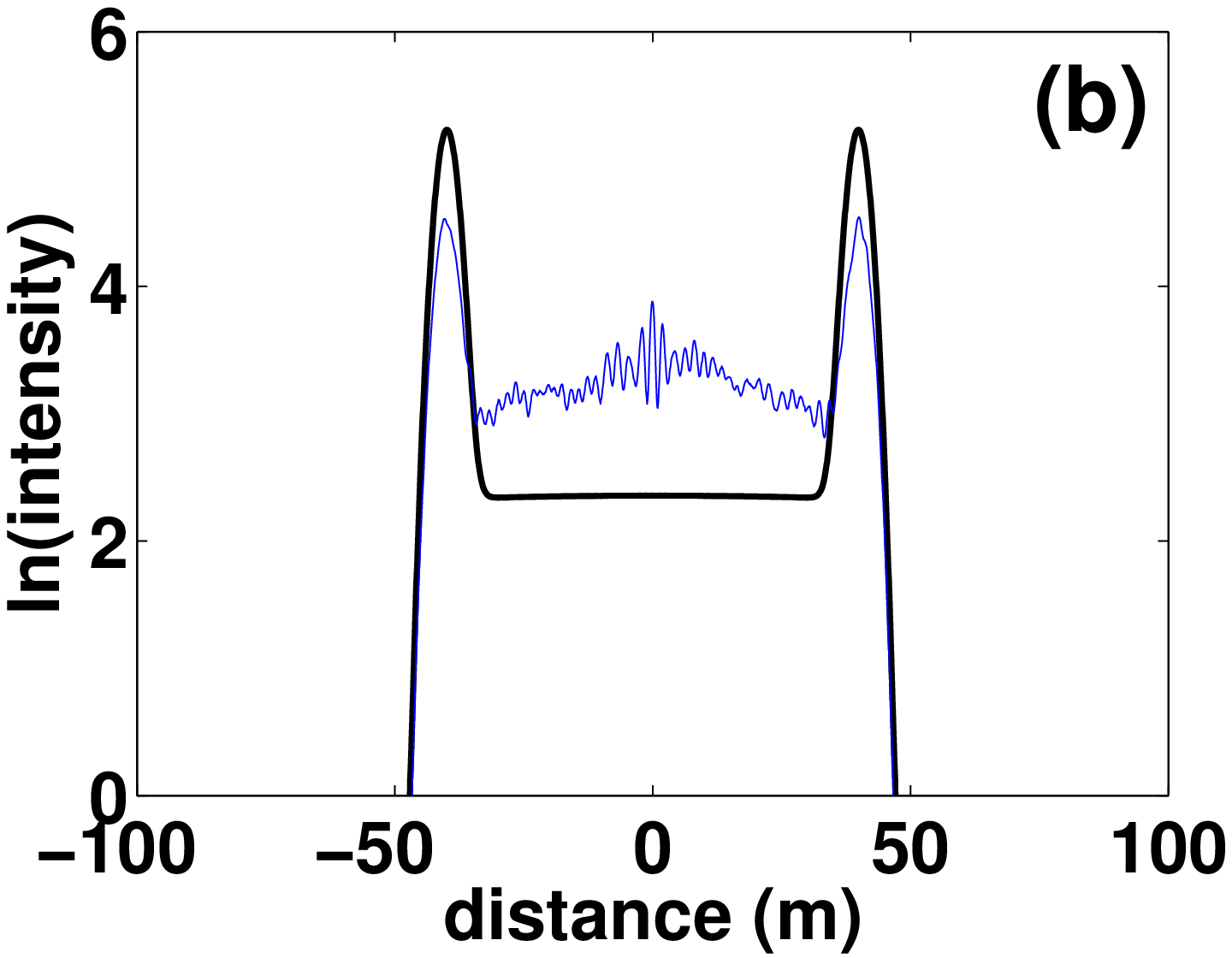}
  \includegraphics[width=7cm]{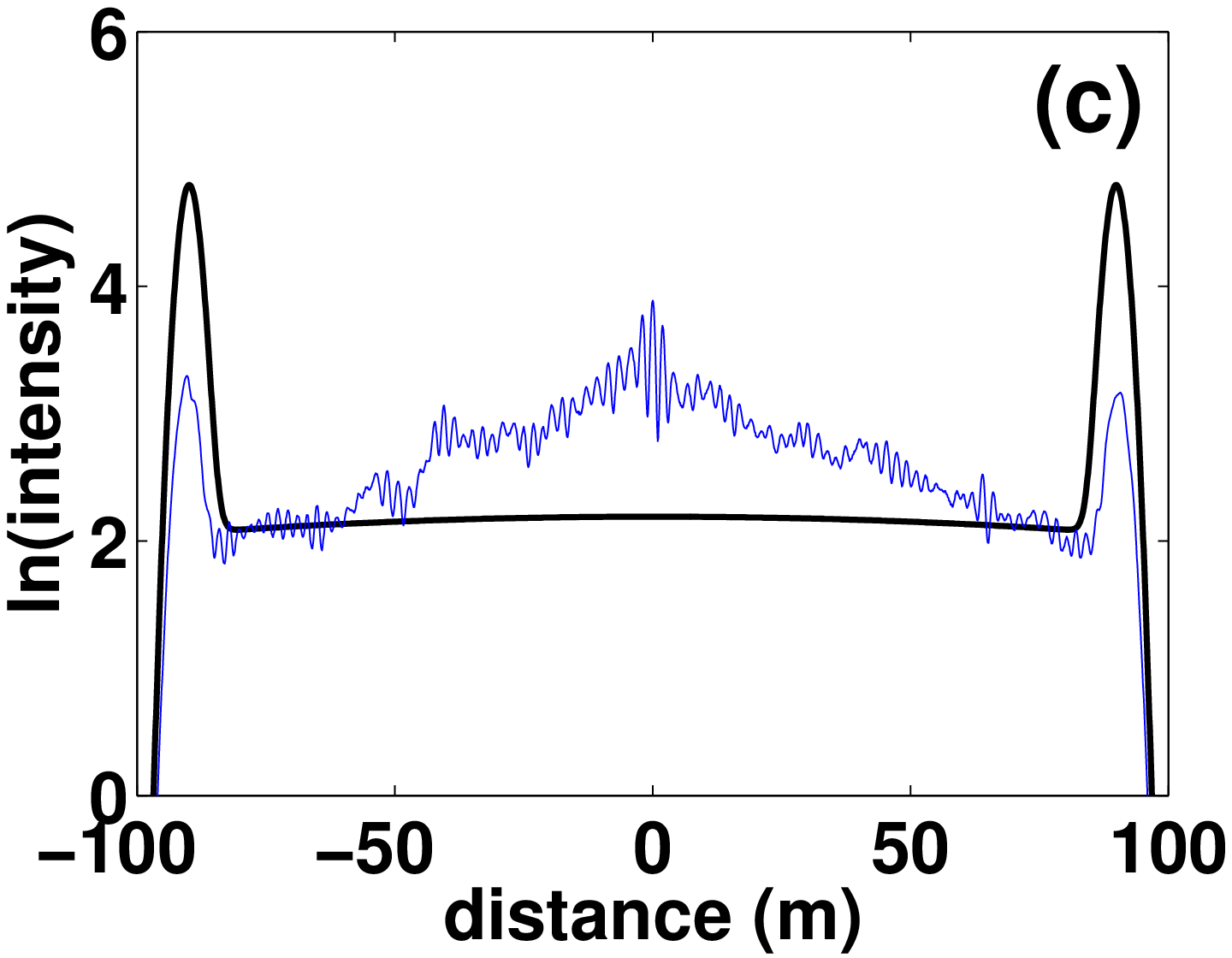}
  \includegraphics[width=7cm]{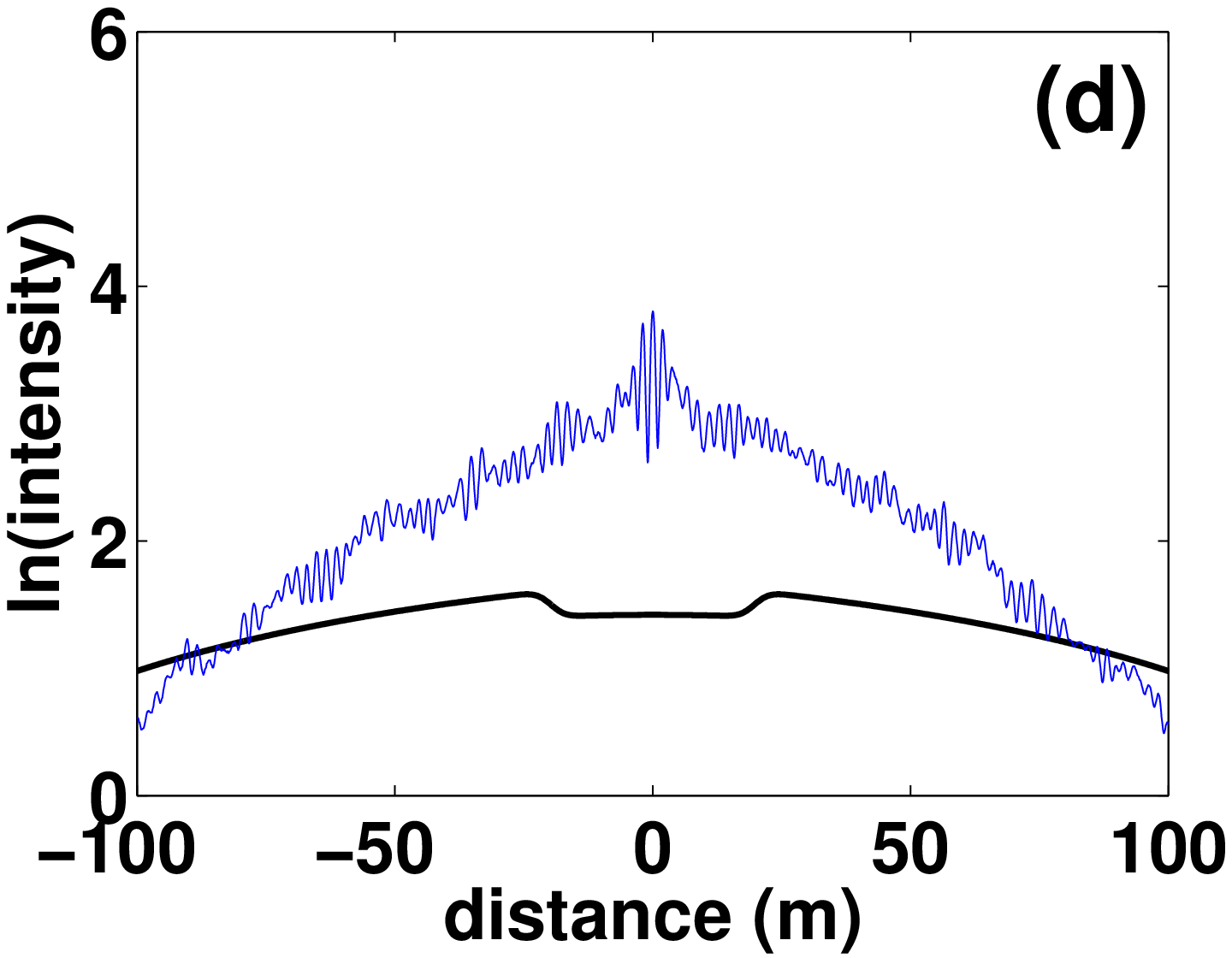}
  \caption{Comparison of numerical results for ensemble-averaged wave
    propagation (thin blue line) and standard KM theory, or RT (thick
    black line). The various panels show: (a) $t = 0$~s, (b) $t =
    0.02$~s, (c) $t = 0.045$~s, (d) $t = 0.11$~s.  The source time
    function is zero-phase and hence acausal (symmetric about $t =
    0$~s).}
  \label{fig:km1}
\end{figure*}
\begin{figure*}
  \includegraphics[width=7cm]{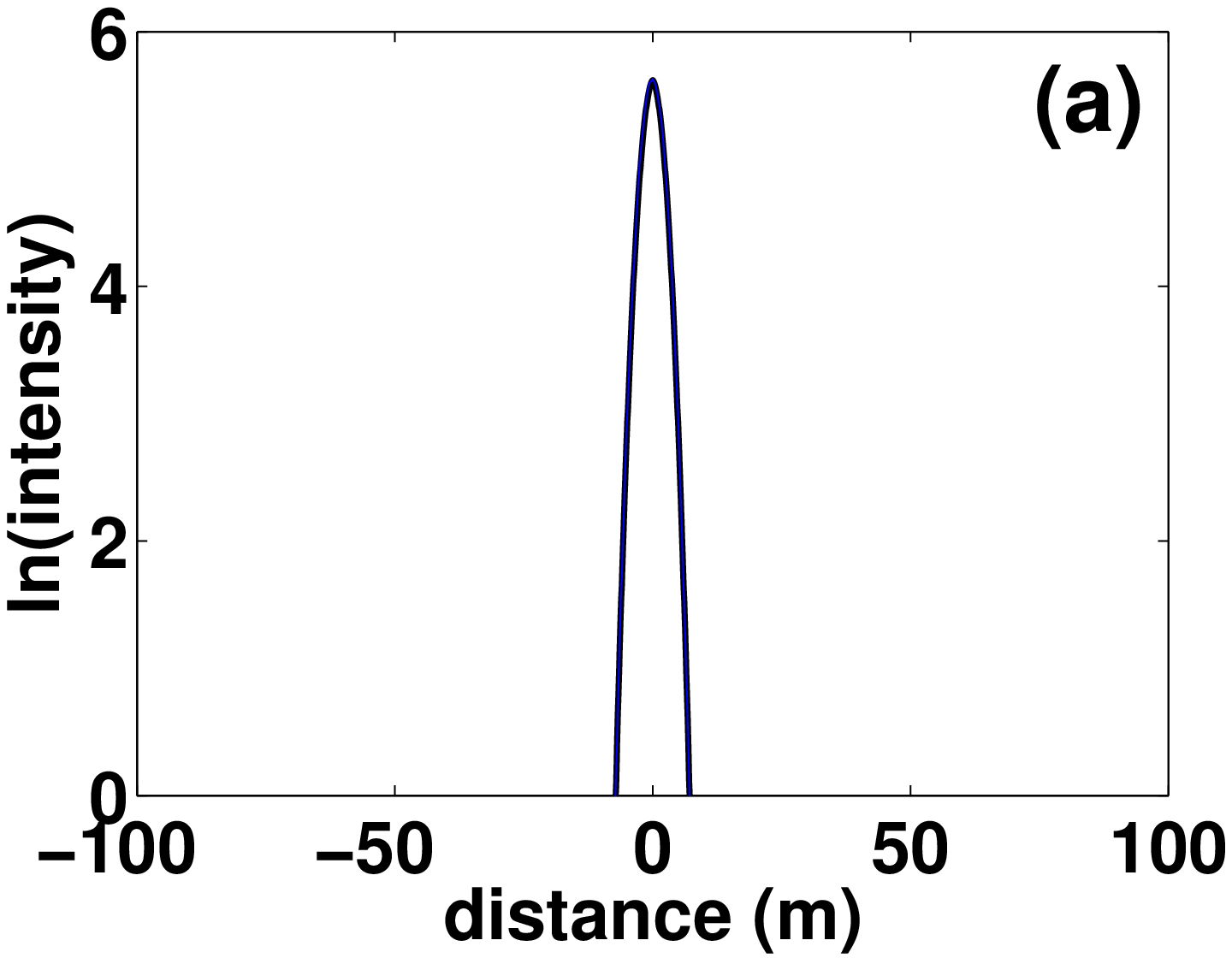}
  \includegraphics[width=7cm]{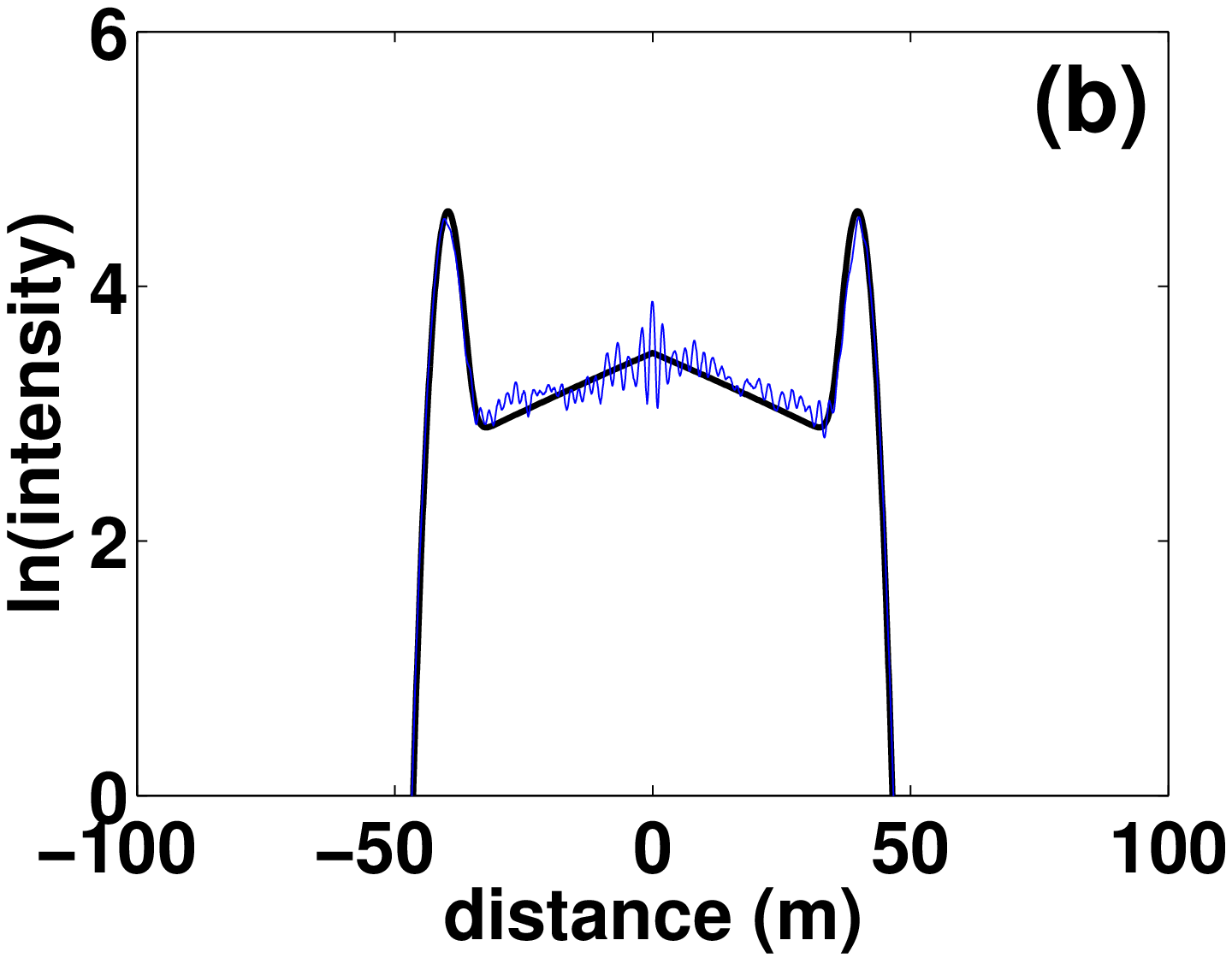}
  \includegraphics[width=7cm]{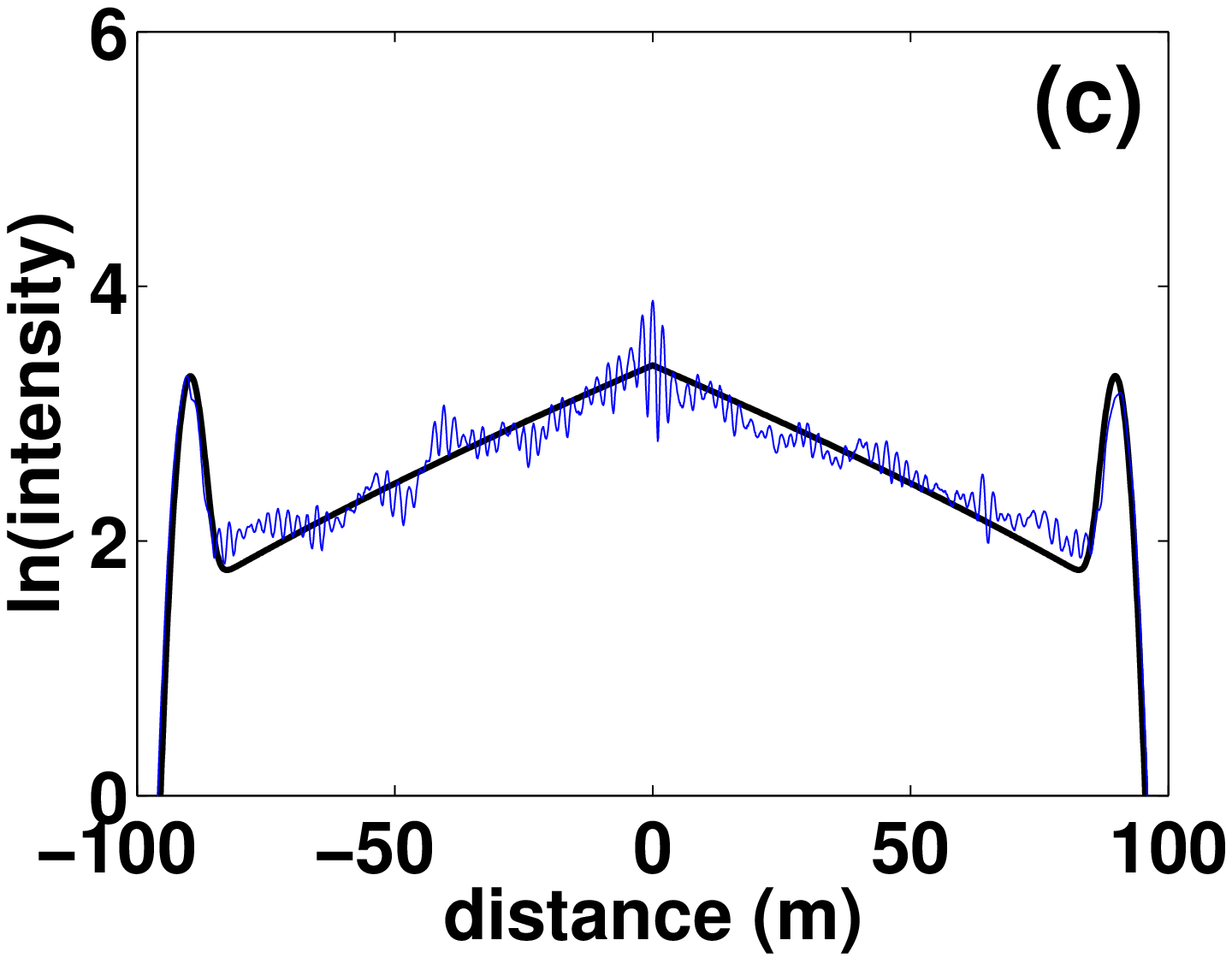}
  \includegraphics[width=7cm]{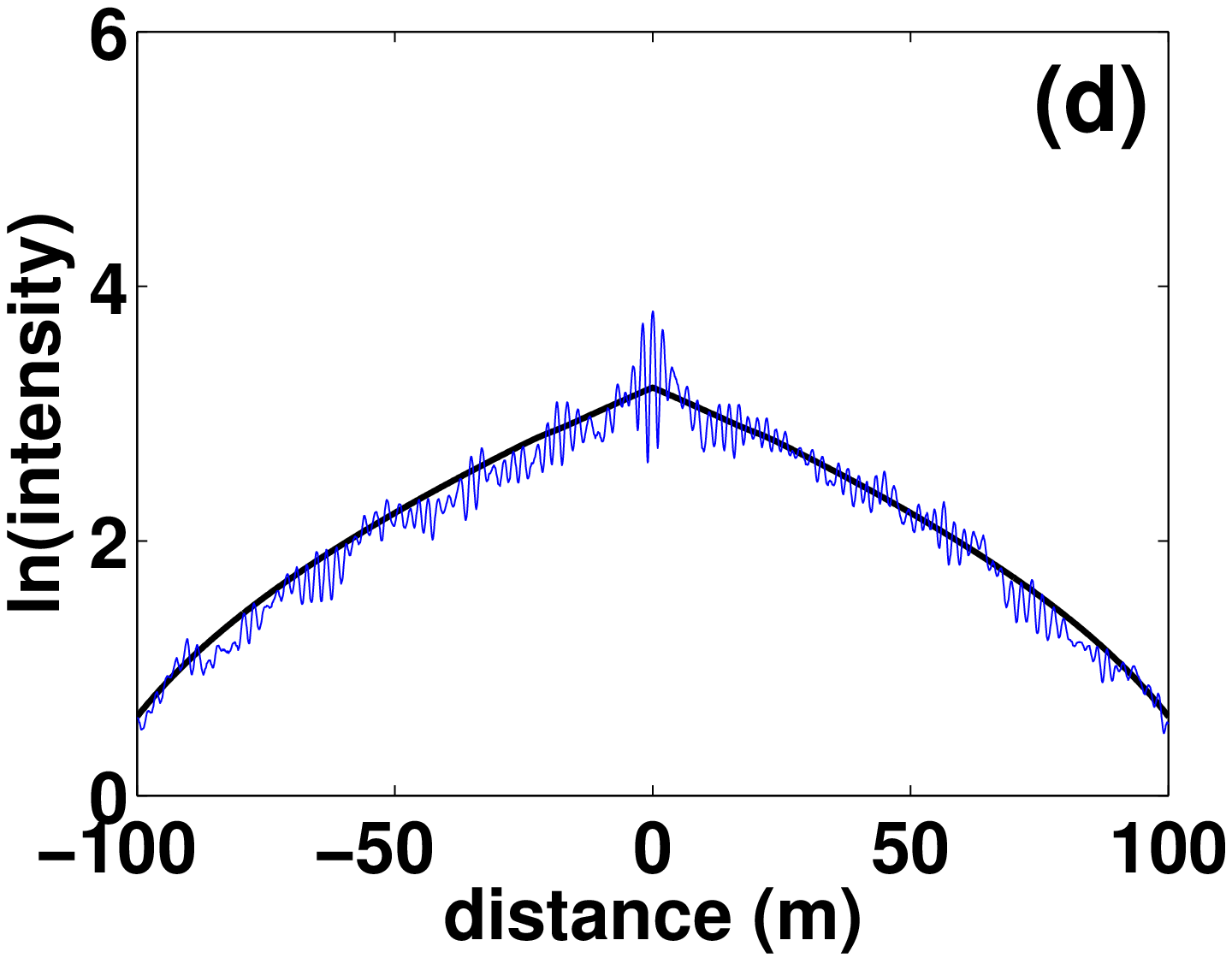}
  \caption{Comparison of numerical results for ensemble-averaged wave
    propagation (thin blue line) and modified KM theory (thick black
    line). The panels show the same times as depicted in
    Figure~\ref{fig:km1}.  The modified KM equations are seen to 
    accurately model the exponential localization of intensity at the
    source position than in Figure~\ref{fig:km1}.}
  \label{fig:km2}
\end{figure*}

\section{Numerical simulations}
Our numerical solution of equations~(\ref{ir}) and (\ref{il}) exploits
staggered grid finite difference methods \cite{virieux}. In this
technique, we calculate the total intensity $I_{t}$ and the intensity
flux $I_{f}$ on different spatial grids that have been shifted by half
of a grid-spacing and on different temporal grids shifted by half of a
time step.  Our purpose is to test whether the modified KM equations,
equations~(\ref{ir}) and (\ref{il}), predict the results of a
wave-based simulation.  Therefore, we also simulate the wave equation
for normally-incident plane waves in a layered medium, excited by a
planar force source:
\begin{equation} 
  {1 \over c^{2}(z)}~{\partial^{2} u \over \partial t^{2}}
  - {\partial^{2} u \over \partial z^{2}} = {1 \over \rho c^{2}(z)} F
  w(t) \delta (z),
  \label{1dwave}
\end{equation}
where $u$ is the displacement field as a function of time $t$ and
spatial coordinate $z$, $c$ is the phase velocity, $\rho$ is the
density of the medium, $F$ is a dimensionless constant related to the
strength of the forcing function, $w(t)$ is the dimensionless source
time function, and $\delta(z)$ is the spatial delta function.  We
simulate equation~(\ref{1dwave}) by the finite-difference method using
centered, second-order approximations for the derivatives.  The
details of the numerical implementation have been previously discussed
in \citet{haney05}.

The setup of our numerical simulation is as follows: for a single
realization, we place 50~scatterers randomly over a depth range of $L
= 200$~m.  The scatterers are lower in propagation velocity (1000 m/s)
than the background medium (2000~m/s) but have the same density.
We excite a source in the center of the 200-m range at
$z_s=0$~m. At the ends of the 200-m range are absorbing boundaries. To
obtain ensemble averages of the total intensity, we first bandpass
filter our numerical results from a single realization with a Gaussian
filter peaked at 500~Hz. We filter in the frequency domain since the
transport properties (i.e., $\ell_{s}$ and $\ell_{loc}$) are strongly
dependent on frequency.  In other words, equations~(\ref{ir}) and
(\ref{il}) model the wave experiment in a particular frequency band.
After filtering, we square the wavefield for each of 100~simulations
with randomly placed scatterers and then add the squared wavefields.
From the ensemble-averaged wavefields, we estimate the extinction mean
free path $\ell_{ext}$ from the decay of the coherent intensity and the
localization length $\ell_{loc}$ from the exponential decay of the
incoherent intensity away from the source. We find $\ell_{ext}=38.1
\pm 0.5$~m and $\ell_{loc}=57.2 \pm 1.7$~m.  These two parameters
enter into equations~(\ref{ir}) and (\ref{il}).  We further find that
the energy velocity of the coherent wave is only slightly altered from
the phase velocity of the background medium (2000~m/s), which is
expected since the scatterers we employ are 1D versions of Rayleigh
scatterers ($B=1/2$, with thickness $d$ much less than the dominant
wavelength) \cite{sheng}, and hence are not resonant scatterers.

The thick black line in Figure~\ref{fig:km1} is the total intensity
from the numerical solution of the standard KM equations (RT,
$\ell_{loc} \rightarrow \infty$), with the wave simulation shown as
the thin blue line. Note that these snapshots are logarithmic in
intensity.  Strong localization effects are evident in the wave
simulation as seen in the sharp exponential peak in the total
intensity at the source position at later times.  This behavior is not
captured in the solution of the standard KM equations, which predict
that the total intensity is {\em flat} around the source position. In
addition, the standard KM equations significantly under-predict the
decay of the coherent wave.  The discrepancy between standard KM
theory and the wave simulation is most evident at $t=0.11$~s in
Figure~\ref{fig:km1}(d), where the wave simulation shows a
concentration of total intensity near the source position.

The simulation for the modified KM equations is the thick black line
in Figure~\ref{fig:km2}. In contrast to the standard KM equations (or
RT), the modified KM equations capture the exponentially-peaked
behavior of the total intensity near the source position and, at all
times, agree well with the wave simulation. Thus, the modified KM
equations are capable of modeling the transport of intensity in 1D
localized media, where interference effects cannot be ignored.  It is
worth emphasizing finally that both the standard KM solution in
Figure~\ref{fig:km1} and the modified KM solution in
Figure~\ref{fig:km2} satisfy global energy conservation.

\section{Conclusion}
With a proper modification to the well known Kubelka-Munk (KM) equations, we are able
to accurately describe the transport of wave intensity in a 1D layered
medium at all times, even when interference effects dominate (e.g.,
wave localization). This is confirmed by numerical simulations comparing
wave simulations and the modified Kubelka-Munk equations. In the
future, we plan to extend our approach, which currently uses only two
fluxes, to a theory valid for 2D and 3D disordered media. One approach to 
this extension would utilize higher dimensional discrete flux theories as
described by \citet{cwilich}.
Such a transport theory will be capable of simultaneously describing the propagating coherent intensity, 
the intensity flux, and the localization transition in 3D.
%Such a theory will have similarities to the SC theory of localization, 
%except that the  

\section{Acknowledgments}
We thank John Scales for many useful discussions. Partially supporting
KvW are the NSF (EAR-0337379) and ARO (DAAD19-03-1-0292).

%\newpage
\bibliography{flux}

%\newpage
\appendix

\section{Verification of The localization length}
\label{app:loca}
Here, we give credence to our use of the term
localization length $\ell_{loc}$ as it appears in
equation~(\ref{eq:I_noxtermsv92}). 
%and thus also in
%equation~(\ref{diff1}). 
By doing so, we justify the expression
for $\ell_{loc}$ in equation~(\ref{scatandloc}). We proceed by finding
the stationary transmission coefficient for a slab geometry in the
case of interference.  We adopt the approach shown by
\citet{vanrossum}, wherein the authors derived the stationary
transmission coefficient $T$ for the case of no interference.  In that
case, $T(L)\approx z_{e}/L$, where $z_{e}$ is the extrapolation length
outside the slab and $L$ is the thickness of the slab
\cite{vanrossum}. In analogy to electronic systems, the behavior $T(L)\approx z_{e}/L$ is an expression 
of Ohm's law.
%Now we examine the same situation but use the steady-state version of equation~(\ref{diff1}), which accounts
%for interference. 

We begin by taking the stationary version of equations~(\ref{ir}) and (\ref{il}) 
%without the presence of any sources
\begin{equation}
  {d I_{f} \over dz} = 
  {\Gamma \over v},
  \label{irs}
\end{equation}
\begin{equation}
  {d I_{t} \over dz} =
  -2\left[{B \over \ell_{s}}+{1 \over \ell_{loc}}\right]I_{f} -
  {\mbox{sgn}(z-z_{s}) \over \ell_{loc}} (I_{t} - |I_{f}|),
  \label{ils}
\end{equation}
where, as discussed before in reference to equations~(\ref{ir}) and (\ref{il}), 
$\Gamma$ is the isotropic (omnidirectional) source term \cite{haney05}.
Let a single stationary (planar) source act at depth $z_{s}$, such that $\Gamma = \delta(z-z_{s})$.
%Therefore, equation~(\ref{ir}) becomes
%\begin{equation}
%  {d I_{f} \over dz} =
%  {\delta(z-z_{s}) \over v}.
%  \label{irs2}
%\end{equation}
%The solution of this simple ODE is
%\begin{equation}
%  I_{f} = { u(z-z_{s}) \over v} + A,
%  \label{irs3}
%\end{equation}
%where $u$ is the step function and $A$ is an integration constant.
Based on physical considerations, we know that $I_{t}$ is symmetric (even) about $z=z_{s}$
and $I_{f}$ is antisymmetric (odd). This, together with the fact that $\Gamma = \delta(z-z_{s})$ in 
equation~(\ref{irs}), leads to the relation 
$I_{f} = \mbox{sgn}(z-z_{s})/2v$ and therefore that
$\mbox{sgn}(I_{f}) = \mbox{sgn}(z-z_{s})$.
%Thus, $I_{f} = 0$ at $z=z_{s}$. With this information, 
%we may solve for the integration constant in equation~(\ref{irs3}).
%Taking $u(0) = 1/2$, we find that $A = -1/2v$. Substituting into equation~(\ref{irs3}) gives
%\begin{equation}
%  I_{f} = { 1 \over 2v}[2u(z-z_{s})-1],
%  \label{irs4}
%\end{equation}
%and, since $\mbox{sgn}(z-z_{s}) = 2u(z-z_{s})-1$, 
%\begin{equation}
%  I_{f} = { 1 \over 2v}\mbox{sgn}(z-z_{s}).
%  \label{irs5}
%\end{equation}
%Furthermore, since the disorder is statistically constant in depth, 
%%From equation~(\ref{irs}), we see that away from the source location, $I_{f}$ is spatially constant ($dI_{f}/dz = 0$). 
%%Thus, there are two possibilities for $\mbox{sgn}(I_{f})$: either $\mbox{sgn}(I_{f}) = \mbox{sgn}(z-z_{s})$ or 
%%$\mbox{sgn}(I_{f}) = -\mbox{sgn}(z-z_{s})$.
%From the last relation, we see that $\mbox{sign}(I_{f}) =
%\mbox{sign}(z-z_{s})$.
%First, note that according to our definition for the intensity flux
%($I_{f}=I_{d}-I_{u}$), we see that $\mbox{sign}(I_{f}) =
%\mbox{sign}(z-z_{s})$.  
Since $\mbox{sgn}(I_{f})|I_{f}| =
I_{f}$, equation~(\ref{ils}) may be written as
\begin{equation}
  {d I_{t} \over dz} =
  -\left[{2B \over \ell_{s}}+{1 \over \ell_{loc}}\right]I_{f} -
  {\mbox{sgn}(z-z_{s}) \over \ell_{loc}} I_{t}.
  \label{il2}
\end{equation}
%In the diffusion approximation, the term $\partial I_{f}/\partial t$
%is neglible, hence equation~(\ref{il2}) becomes
%\begin{equation}
%  {\partial I_{t} \over \partial z} \approx
%  -\left[{2B \over \ell_{s}}+{1 \over \ell_{loc}}\right]I_{f} -
%  {\mbox{sgn}(z-z_{s}) \over \ell_{loc}} I_{t}.
%  \label{il3}
%\end{equation}
Solving this equation for $I_{f}$ allows a substitution for $I_{f}$ in
equation~(\ref{irs}).
%Introducing the transport mean free path $\ell_{tr} = \ell_{s}/2B$ \cite{haney05}, this gives
At depths away from the source (for $z \neq z_{s}$), this gives an expression in terms of $I_{t}$ only
\begin{equation}
  {d^{2} I_{t} \over dz^{2}} + {1 \over \ell_{loc}}{d \over dz} \left[\mbox{sgn}(z-z_{s}) I_{t}\right] = 0.
  \label{diff1}
\end{equation}
%Note that assuming the other case, $\mbox{sign}(I_{f}) = -\mbox{sgn}(z-z_{s})$, solving for $I_{f}$, and substituting 
%for $I_{f}$ in equation~(\ref{ir_stat2}) leads to the 
%same form for equation~(\ref{diff1}). Thus, equation~(\ref{diff1}) does not depend on whether 
%$\mbox{sgn}(I_{f}) = \mbox{sgn}(z-z_{s})$ or $\mbox{sgn}(I_{f}) = -\mbox{sgn}(z-z_{s})$.
For the case of no interference, $\ell_{loc} \rightarrow \infty$, 
equation~(\ref{diff1}) reduces to the Laplace equation (the diffusion equation in
the stationary case). However, when interference is taken into
account, the equation is a modified Laplace equation.

In preparation for an application of the standard approach shown by
\citet{vanrossum}, we proceed by investigating how the extrapolation length outside of a slab of thickness $L$
changes when interference is accounted for. We use the well-known
approach of \citet{morse} to define the extrapolation length.
Consider a slab containing randomly located thin layers extending from
$z=0$ to $z=L$. Outside of this interval, the medium is homogeneous. Take  
a planar source of intensity at some $z_{s}<0$, outside of the slab.  
Since the entire slab extends over
$z=0$ to $z=L$, we have $z_{s} < z $ and thus $\mbox{sgn}(z-z_{s})
= 1$ for all points $z$ within the slab. In this case, equations~(\ref{il2}) and (\ref{diff1}) are, for all points $z$ inside of the slab, given by
\begin{equation}
  {d I_{t} \over dz} =
  -\left[{2B \over \ell_{s}}+{1 \over \ell_{loc}}\right]I_{f} -
  {I_{t} \over \ell_{loc}},
  \label{il22}
\end{equation}
and
\begin{equation}
  {\partial^{2} I_{t} \over \partial z^{2}} + {1 \over \ell_{loc}}{\partial I_{t} \over \partial z} = 0.
  \label{diff2}
\end{equation}

At the far end of the slab $z=L$, we require there be no up-going
intensity $I_{u} = (I_{t}-I_{f})/2 = 0$.  Now using
equation~(\ref{il22}), we substitute for $I_{f}$ in the relation
$I_{t}-I_{f} = 0$, giving an equation in terms of $I_{t}$ only
\begin{equation}
  I_{t} + {\ell_{tr}\ell_{loc} \over \ell_{loc} + \ell_{tr}} \left[{d I_{t} \over dz} + {I_{t} \over \ell_{loc}}\right] = 0,
  \label{extrap3}
\end{equation}
where we use the transport mean free path $\ell_{tr} = \ell_{s}/2B$ to
make the notation concise \cite{haney05}.
%Note that this expression becomes equation~(\ref{extrap1}) in the limit of no interference, $\ell_{loc} \rightarrow \infty$.
%Note that, since at $z=L$ we are considering $z>z_{s}$, the signum
%term in equation~(\ref{il2}) is equal to unity, $\mbox{sgn}(z-z_{s}) =
%1$.  
We can rewrite equation~(\ref{extrap3}) as
\begin{equation}
  I_{t} + \alpha {d I_{t} \over dz} = 0.
  \label{extrap4}
\end{equation}
%By comparison with equation~(\ref{extrap1}), we can see that the extrapolation length in the case of
%interference is different than without interference - it is $z_{e} = \ell_{tr}\ell_{loc}/(\ell_{loc} + 2\ell_{tr})$.
where $\alpha = \ell_{tr}\ell_{loc}/(\ell_{loc} + 2\ell_{tr})$.
Equation~(\ref{extrap4}) means that, near $z=L$, $I_{t}\approx
C(L+\alpha-z)/\alpha$ where $C$ is a dimensioning constant.
Within this approximation, $I_{t}= 0$ at $z = L+\alpha$; therefore,
the extrapolation length $z_{e}$ - the distance outside of the slab
where $I_{t}$ vanishes - is equal to $\alpha$. That is, $z_{e} =
\ell_{tr}\ell_{loc}/(\ell_{loc} + 2\ell_{tr})$.  One can see in this
expression that, for no interference ($\ell_{loc} \rightarrow
\infty$), $z_{e} = \ell_{tr}$ which is the usual extrapolation length
encountered in 1D when interference is neglected \cite{haney05}.

At the side of the slab on which the source of intensity is incident,
at $z=0$, we require the down-going intensity to be equal to the
incident intensity, $I_{0}$.  Thus, $I_{d} = (I_{t}+I_{f})/2 = I_{0}$
at $z=0$. Using equation~(\ref{il22}), we substitute for $I_{f}$ in the
relation $(I_{t}+I_{f})/2 = I_{0}$, giving
\begin{equation}
  I_{t} - {\ell_{tr}\ell_{loc} \over \ell_{loc} + \ell_{tr}} \left[{\partial I_{t} \over \partial z} + {I_{t} \over \ell_{loc}}\right] = 2I_{0}.
  \label{extrap5}
\end{equation}
%Note that, since we are considering $z=0$ and the source of intensity
%comes from $z_{s} < 0 $, the signum term in equation~(\ref{il2}) is
%again equal to unity, $\mbox{sgn}(z-z_{s}) = 1$.
Equation~(\ref{extrap5}) may be rewritten as
\begin{equation}
  I_{t} - \ell_{tr} {\partial I_{t} \over \partial z} = {2I_{0}(\ell_{loc} + \ell_{tr}) \over \ell_{loc}}.
  \label{extrap6}
\end{equation}
% By comparison with equation~(\ref{extrap2}), we see that near $z=0$,
Near $z=0$, the solution is approximately given by
\begin{equation}
  I_{t} \approx {2I_{0}(\ell_{loc} + \ell_{tr}) \over \ell_{loc}} \left[{z \over \ell_{tr}} + 2\right].
  \label{extrap7}
\end{equation}
Within this approximation, at a distance equal to the (interference
adjusted) extrapolation length outside of the slab, $z=-z_{e}$,
$I_{t}$ is therefore given by
\begin{equation}
  I_{t} = 2I_{0} {(\ell_{loc} + \ell_{tr}) (\ell_{loc} + 4\ell_{tr}) \over \ell_{loc}(\ell_{loc} + 2\ell_{tr})} = 2I_{0} \Delta,
  \label{extrap8}
\end{equation}
where we represent the term containing $\ell_{loc}$ and $\ell_{tr}$ by $\Delta$.

% In the steady-state case, the modified diffusion equation, equation~(\ref{diff1}), is (without forcing terms)
%As noted in the previous paragraph, for the entire slab extending over
%$z=0$ to $z=L$ we have $z_{s} < z $.  Therefore, $\mbox{sgn}(z-z_{s})
%= 1$ for all $z$ within the slab and equation~(\ref{diff1}) is given
%by
%\begin{equation}
%  {\partial^{2} I_{t} \over \partial z^{2}} + {1 \over \ell_{loc}}{\partial I_{t} \over \partial z} = 0.
%  \label{diff2}
%\end{equation}
% Again, note that for the current case $\mbox{sgn}(z-z_{s}) = 1$ in equation~(\ref{diff1}).
Following the method employed by \citet{vanrossum}, we seek to solve
equation~(\ref{diff2}) with the boundary conditions $I_{t} = 0$ and
$I_{t} = 2I_{0}\Delta$ at the (interference adjusted) extrapolation
lengths, $z=L+z_{e}$ and $z=-z_{e}$ respectively.  The solution is
\begin{equation}
  I_{t}(z) = 2I_{0}\Delta \left[{ e^{-z/\ell_{loc}} - e^{-(L+z_{e})/\ell_{loc}} \over e^{z_{e}/\ell_{loc}} - e^{-(L+z_{e})/\ell_{loc}} }\right].
  \label{diffsol2}
\end{equation}
The steady-state transmission coefficient, $T(L) = I_{t}(z=L)/I_{0}$,
is
%\begin{equation}
\begin{eqnarray}
  T(L) = 2\Delta e^{-L/\ell_{loc}} \left[{1 - e^{-z_{e}/\ell_{loc}} \over e^{z_{e}/\ell_{loc}} - e^{-(L+z_{e})/\ell_{loc}} }\right] \approx \nonumber \\
  2\Delta e^{-L/\ell_{loc}} \left[{1 - e^{-z_{e}/\ell_{loc}} \over e^{z_{e}/\ell_{loc}} }\right].
  \label{trans_int}
%\end{equation}
\end{eqnarray}
where the approximation is for $z_{e} \ll
L$. %In contrast to equation~(\ref{trans_noint}),
This expression shows that when interference is taken into account the
steady-state transmission coefficient goes down {\it exponentially} as
a function of the length of the slab $L$ - a hallmark of localization in the stationary case. 
This behavior is in stark contrast to
$T(L)\approx z_{e}/L$ \cite{vanrossum} obtained when interference is
neglected ($\ell_{loc} \rightarrow \infty$). The fact that the length
scale controlling the exponential decay with $L$ in
equation~(\ref{trans_int}) is $\ell_{loc}$ supports the use of this
term in equation~(\ref{eq:I_noxtermsv92}) and, as a consequence, the
expression for $\ell_{loc}$ in equation~(\ref{scatandloc}).
% This is the hallmark of a fully localized system \cite{wiersma97},
% one that no longer conducts (within an exponentially vanishing
% amount).

\end{document}